\documentstyle[aps,prl,twocolumn,epsf,epsfig]{revtex}
\def\beq{\begin{equation}}
\def\eeq{\end{equation}}
\def\beqarr{\begin{eqnarray}}
\def\eeqarr{\end{eqnarray}}

\begin{document}
\draft
\twocolumn[\hsize\textwidth\columnwidth\hsize\csname @twocolumnfalse\endcsname

\title{Magnetic Percolation and the Phase Diagram
of the Disordered RKKY Model}
\author{D. J. Priour, Jr and S. Das Sarma}
\address{Condensed Matter Theory Center, Department of Physics,
University of Maryland, College Park, MD 20742-4111}

\date{\today}

\maketitle

\begin{abstract}
     We consider ferromagnetism in spatially randomly located 
magnetic moments, as in a diluted  
magnetic semiconductor, coupled
via the carrier-mediated indirect exchange RKKY interaction.  We obtain via
Monte Carlo the magnetic phase diagram as a function of the impurity moment 
density $n_{i}$ and the relative carrier concentration
$n_{c}/n_{i}$.  As evidenced by the diverging correlation 
length and magnetic susceptibility, the boundary between 
ferromagnetic (FM) and non-ferromagnetic (NF) phases constitutes
a line of zero temperature critical points which can be 
viewed as a magnetic percolation transition.  In the dilute
limit, we find that bulk ferromagnetism vanishes for $n_{c}/n_{i}>.1$.
We also incorporate the local antiferromagnetic direct 
superexchange interaction between nearest neighbor impurities, and
examine the impact of 
a damping factor in the RKKY range function.  
\end{abstract}

%\pacs{}
\pacs{PACS numbers: 75.50.Pp,75.10.-b,75.10.Nr,75.30.Hx}
%\vspace{-.5cm}
]

There has been substantial recent interest in the old problem
of long-range ferromagnetic ordering in localized ``impurity''
magnetic moments induced by indirect exchange (``RKKY'') interaction
mediated through (effectively) ``free'' carriers (either 
electrons or holes).  The recent interest arises from the context
of ferromagnetic ordering in diluted magnetic semiconductors (DMS),
such as $\textrm{Ga}_{1-x}\textrm{Mn}_{x}\textrm{As}$ with 
$x \approx 0.1$, where the Mn dopants act both as the impurity
magnetic moments and as acceptors producing the carriers 
(which happen to be holes for GaMnAs) mediating the RKKY 
coupling.  The standard model for DMS ferromagnetism has been the 
carrier-mediated RKKY interaction, and understanding the RKKY
magnetic phase diagram, therefore, takes on particular 
significance.  This is important in view of the diluted and the 
random nature of the spatial distribution of the impurity 
magnetic moments which could lead to substantial frustration in 
the magnetic interaction between the impurity moments due to
oscillatory nature of the RKKY coupling.  The latter
yield substantial antiferromagnetic (AF) couplings
which have the potential to disrupt ferromagnetism.  Our goal in this 
paper is to numerically obtain the RKKY magnetic phase diagram
in a disordered DMS system via direct Monte Carlo simulations.

The complex interplay of the
the long-ranged oscillatory
behavior of the RKKY interaction and strong 
disorder makes simple theoretical statements 
difficult, and it is not 
obvious a priori whether in the dilute limit the 
ferromagnetic state is supported for any choice
of system parameters even at $T=0$.   
Mean field treatments such as the Curie Weiss continuum
mean field theory (MFT) 
are problematic, because they fail to 
take into account the discrete crystal lattice 
and thermal fluctuations and assume a ferromagnetic
ground state without providing any means of assessing 
the validity of this assumption.  
In this Letter, we rigorously take into account  
positional disorder and we obtain the true
ground state spin configurations, finding
that the RKKY model \textit{does}
support a ferromagnetic phase, albeit only for a limited
parameter range.  In addition, we demonstrate via 
direct Monte Carlo that the 
transition from the non-ferromagnetic (NF) to the ferromagnetic (FM)
phase is marked by the percolation of magnetic 
clusters which grow in size as AF couplings are reduced and
ultimately coalesce to yield long-range ferromagnetic
order; this constitutes zero temperature percolation critical behavior.  
We neglect all quantum fluctuations, but our interest being the 
interplay of disorder and magnetic interaction, our results should 
in general be valid with respect to the existence (or not) of the 
FM phase.

Although the specific
calculations described in this work are motivated
by possible carrier-mediated RKKY 
ferromagnetism in DMS systems, 
our results also apply more broadly 
to a variety of disordered magnets with competing
interactions (e.g. the Edwards-Anderson model which we 
have also examined and found similar behavior) 
where the NF to FM transition at (T=0) occurs
via magnetic percolation.  Note that we do not 
attempt to characterize the NF phase (which may be
a simple paramagnet or a subtle glassy phase) except to 
emphasize that it does not have long-range FM 
ordering.

Our model physical system is
$\textrm{Ga}_{1-x}\textrm{Mn}_{x}\textrm{As}$ (by far the 
most studied DMS system) where $x_{i} \approx 0.01-0.1$, the 
concentration of Mn dopants, is in the dilute limit.  We
assume that Mn impurities only occupy Ga sites in the zinc-blende
GaAs (fcc) lattice with a lattice constant $a$.
The large spin ($S = 5/2$) of the Mn moments permits a classical
treatment of the spins, which we regard as Heisenberg spins
governed by the Hamiltonian ${\mathcal H} = 
\sum_{ij}J(r_{ij}){\bf S}_{i} \cdot {\bf S}_{j} + 
\sum_{\langle ij \rangle}J^{\mathrm{AF}}{\bf S}_{i} \cdot {\bf S}_{j}$, 
where in the 
first term $r_{ij}$
is the separation between moments $i$ and $j$ and $J(r)$ is the 
RKKY range function given in three dimensions by
$J(r) = J_{0}e^{-r/l}r^{-4}[\sin(2k_{\mathrm{F}}r)-2k_{\mathrm{F}}r
\cos(2k_{\mathrm{F}}r)]$; $k_{\mathrm{F}} = (\frac{3}{2} \pi^{2} n_{c})^{1/3}$
is the Fermi wave number, and $n_{c}$ is the hole density (a 
closely related quantity is $x_{c} = n_{c}/4$, the number of 
holes per lattice site).
$J_{0}(>0)$ is related to the local Zener coupling $J_{pd}$
between the Mn local moments and the hole spins.  In particular,
we have $J_{0} \propto mJ_{pd}^{2}$ with $m$ being the hole 
effective mass and the $S^{2} = (5/2)^{2}$ factor being 
absorbed into $J_{0}$.  With DMS materials being at best 
poor metals, we also introduce in the RKKY range function a 
damping factor $e^{-r/l}$ where the damping scale $l$ is related to the 
carrier mean free path or perhaps a carrier localization 
length arising from Anderson localization of the 
free carriers; however, we initially set $l= \infty$ in order
to study the long-ranged RKKY coupling.  For the latter, 
an important subtle question is whether the full RKKY model, 
with its long-ranged oscillatory behavior,
supports a ferromagnetic ground state for strongly disordered
systems.
In the second term of ${\mathcal H}$,
$J^{\mathrm{AF}}$ is the local antiferromagnetic superexchange 
coupling which is relevant only for neighboring impurities on the fcc 
lattice.  We note that very large values of $J_{0}$ (which
tend to localize the hole carriers and hamper indirect 
exchange~\cite{tres}) are also 
deleterious to ferromagnetism and lower $T_{c}$, but this effect is not 
examined in this paper.  We also ignore all band structure effects, 
which should be adequate for qualitative purposes.  Our goal
here is to obtain explicitly the ideal RKKY phase diagram, 
rather than calculate results for a specific material
or compare with experiments.

Salient length scales  
include the mean
spacing between impurities, $l_{s} \equiv n_{i}^{-1/3}$ and 
the scale of the oscillations in the RKKY
range function, $k_{\mathrm{F}}^{-1}$.  
Hence $k_{\mathrm{F}}l_{s}$ determines the
relative importance of ferromagnetic and antiferromagnetic couplings;
for $k_{\mathrm{F}}l_{s} \sim 1$ RKKY oscillations tend to disrupt 
ferromagnetic order yielding a NF ground state, while one expects
ferromagnetic interactions to dominate for
$k_{\mathrm{F}}l_{s} \ll 1$.  Since $k_{\mathrm{F}}l_{s} \propto
(n_{c}/n_{i})^{1/3}$, and due to its experimental relevance, $n_{c}/n_{i}$ is
a useful parameter of merit for gauging the importance of AF 
interactions.  While we concentrate on ferromagnetic ordering
for $n_{c}/n_{i} \ll 1$ regime, 
other work has examined the $n_{c}/n_{i} \gg 1$ limit, deep in 
the NF phase~\cite{cuatro,cinco}.
We will see that, although a stable ferromagnetic
phase is supported by the pure RKKY model, 
it occurs only for a relatively narrow $n_{c}/n_{i}$
domain; in particular, in the dilute limit, 
we find long-range ferromagnetic 
order at $T = 0$ only for $n_{c}/n_{i} < .1$.  
    Previous Monte Carlo calculations~\cite{cincoii,cincoiii,cincoiv}
    have examined on
a qualitative basis the impact of competing interactions 
on the ferromagnetic state.  We show explicitly that 
the transition of the ferromagnetic to the non-ferromagnetic
phase at $T=0$ is a disorder driven critical transition 
involving magnetic percolation.

    In the presence of strong disorder, $n_{c}/n_{i}$ has a role 
very similar to temperature ($k_{\mathrm{B}}T/J_{0}$);
just as thermal fluctuations at 
higher temperatures disrupt  
ferromagnetic order by flipping spins, a larger $n_{c}/n_{i}$ 
is associated with strong AF interactions which prevent many pairs 
of spins from aligning.
    For large enough $n_{c}/n_{i}$,
only spins in close proximity are ferromagnetically correlated.
As $n_{c}/n_{i}$ is decreased,
these small correlated groups of spins increase in size.  Ultimately, 
the magnetic clusters span the entire system and magnetic percolation occurs,
signaling the appearance of long-range ferromagnetic order.
This is the extended free-carrier analog of the bound magnetic
polaron percolation ferromagnetic transition recently 
discussed~\cite{cincovi} in the context of strongly localized DMS materials.
To identify magnetic percolation and thereby locate the NF/FM
phase boundary, we determine the typical size of the magnetic 
clusters using a standard technique to calculate the
ferromagnetic correlation length $\xi$ within Monte Carlo~\cite{seis}.
Magnetic percolation and concomitant long-range ferromagnetism
occurs when $\xi$ becomes comparable to the system size $L$.

In our Monte 
Carlo calculations, we average over at least 500 disorder 
realizations.  We obtain ground state configurations via Monte Carlo simulated 
annealing; thermal fluctuations are provided by the Heat Bath 
technique~\cite{siete}. 
To exploit finite size scaling, we examine the behavior of the  
normalized correlation length $\xi/L$ (where $L$ 
is the linear dimension of the system) as a function of 
system size.    
In the NF phase, $\xi/L$ diminishes
as $L$ is increased, while for ferromagnetic order,
$\xi/L$ increases with increasing $L$.  One seeks 
the FM/NF phase boundary where $\xi/L$ is 
constant in $L$ for the critical value of $n_{c}/n_{i}$.  
To minimize finite size effects, we
examine a range of system sizes where the mean number of 
spins $\langle N \rangle$ contained in the system is at 
least on the order of 1000.
To calculate the Curie Temperature $T_{c}$, we also 
examine $\xi/L$, 
but we vary the temperature $T$ instead 
of $n_{c}/n_{i}$.  Again, in obtaining $T_{c}$, we seek the 
critical $\xi/L$ curve.

\begin{figure}
\centerline{\psfig{figure=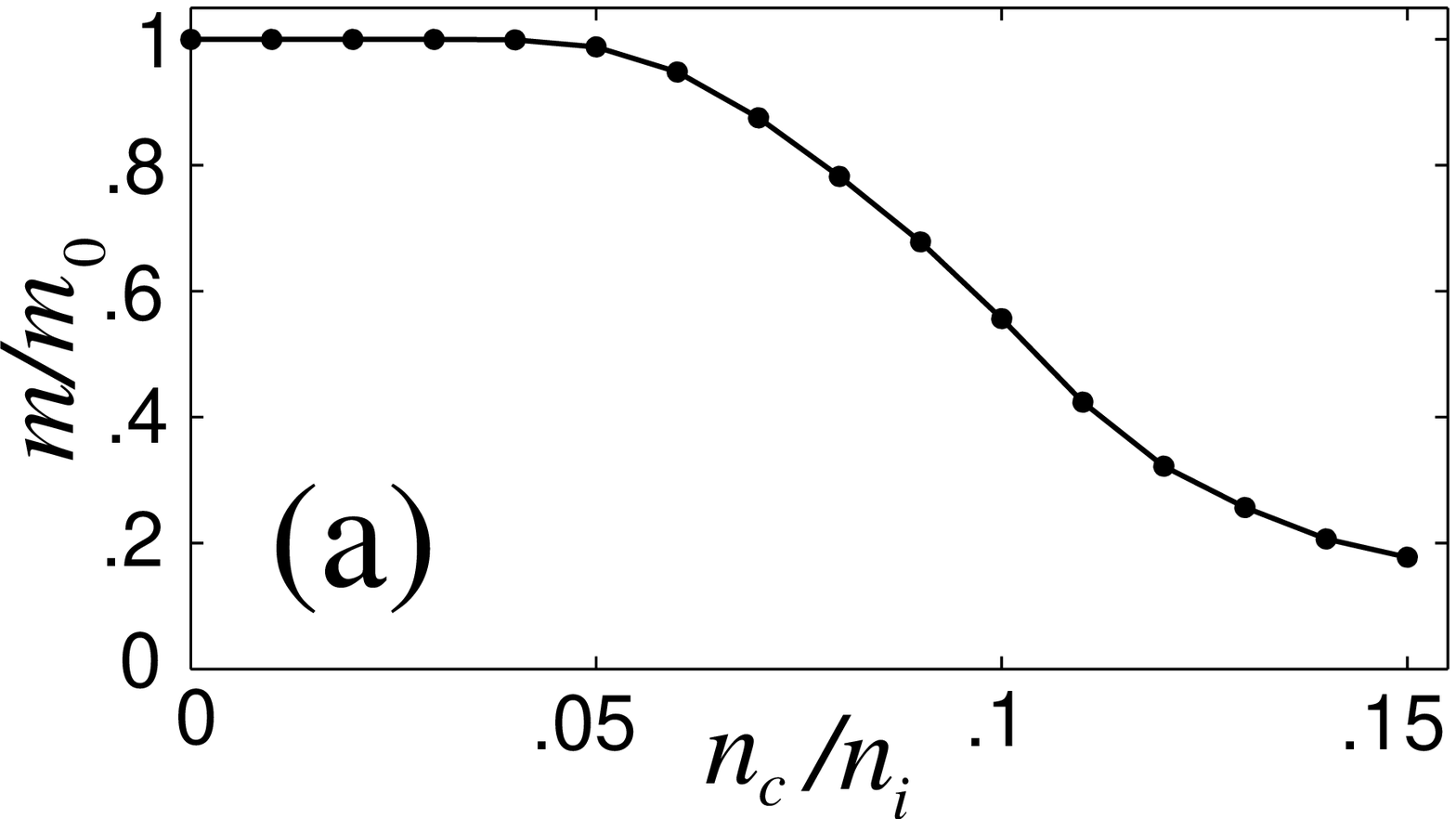,width=2.3in}}
\vspace{.2cm}
\centerline{\psfig{figure=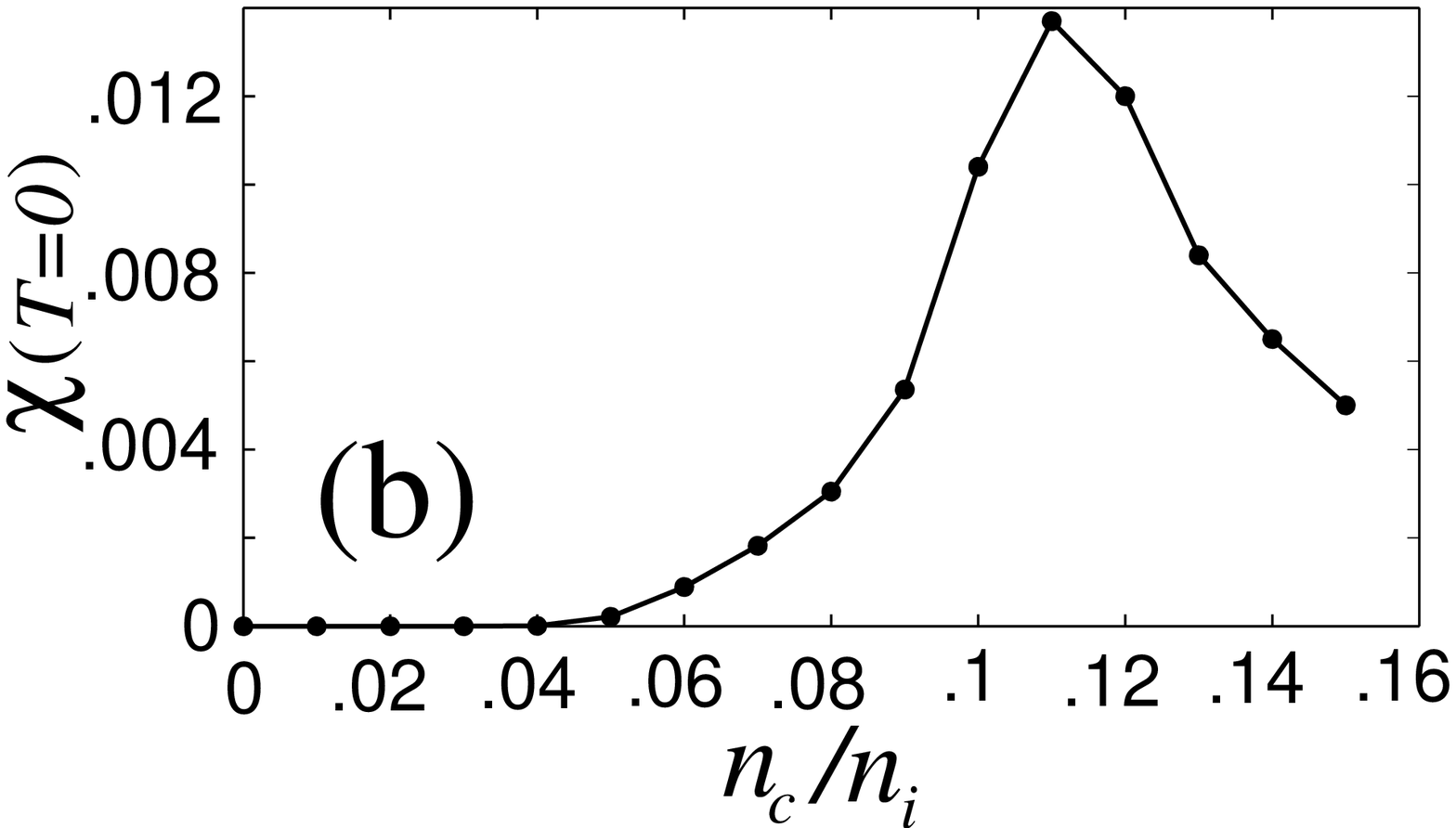,width=2.5in}}
\vspace{.2cm}
\centerline{\psfig{figure=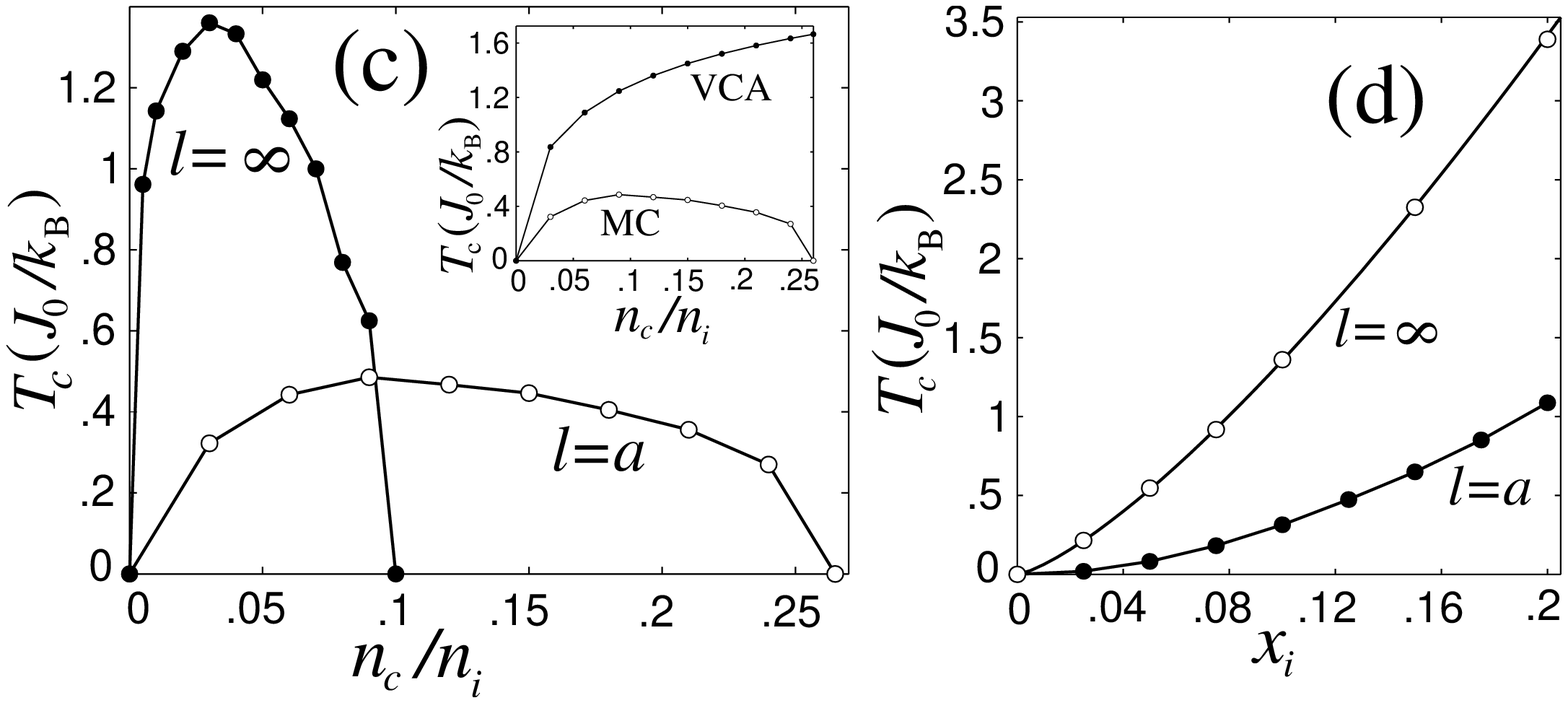,width=3.2in}}
\caption{$T=0$ magnetization and susceptibility,
and $T_{c}$ plots versus $n_{c}/n_{i}$ for $x_{i} = 0.1$;
results are shown for the undamped RKKY with no
AF superexchange coupling.  Panels (a) and (b) display
the zero temperature magnetization and susceptibility,
respectively for $\langle N \rangle = 1000$.  Panel (c)
depicts Curie temperatures versus $n_{c}/n_{i}$.  The
filled circles correspond to the undamped ($l=\infty$) case
and the open circles to the damped ($l=a$) RKKY model.
The inset of (c) shows VCA and MC $T_{c}$'s for $l=a$.
Panel (d)
is a graph of $T_{c}$ versus $x_{i}$ with $n_{c}/n_{i} = 0.03$
for $l=\infty$ and $l=a$.  For $l=\infty$, open circles are
Monte Carlo results while the
solid line in (d) is a theoretical curve, given by
$29.2x_{i}^{4/3}$; for $l=a$, closed circles are Monte Carlo $T_{c}$'s,
and the solid line is a guide to the eye.}
\label{Fig:fig1}
\end{figure}

     In Fig.~\ref{Fig:fig1}, we display calculated characteristics
which would be accessible in 
experiment; these results are obtained for a large system with 
$\langle N \rangle \sim 1000$.  Panel (a) of Fig.~\ref{Fig:fig1}
is a graph of the ferromagnetic order parameter $m = \left[
\langle \left| {\bf S} \right| \rangle \right]$ (square
brackets indicated disorder averaging), the magnetization
obtained for the undamped ($l=\infty$) case.
For small $n_{c}/n_{i}$, one sees what appears to be a
plateau where spins are essentially perfectly collinear;
a reduction in the polarization begins for larger $n_{c}/n_{i}$
with the magnetization becoming strongly attenuated for $n_{c}/n_{i} \sim
0.1$.  It is tempting to regard the ``plateau'' and noncollinear
regimes as indicative of qualitatively distinct phases, but this 
notion can be seen to be illusory if one considers that
for any spin, there is always 
a finite probability 
of having a void large enough that the distance from the 
spin to its nearest neighbor is greater than the distance to 
the first zero of the RKKY oscillation.  
Since in this situation 
the moment would not interact 
ferromagnetically with its nearest neighbor, it is reasonable to 
assume that there is 
always some noncollinearity for 
any finite value of $n_{c}/n_{i}$.  
%%%%%%%%%%%%%%%%%%%%%% FIG. 1 %%%%%%%%%%%
%%%%%%%%%%%%%%%%%%%%%%%%%%%%%%%%%%%%%%%%%
 
On the other hand, for larger $n_{c}/n_{i}$ it 
is also not evident in the graph of the magnetization in 
Fig.~\ref{Fig:fig1} (a) where the 
polarization drops to zero (as would happen in the 
thermodynamic limit), and one has instead a long tail for
$n_{c}/n_{i} \gtrsim 0.1$.  This slow decay of the magnetization for
even fairly large systems (e.g. $\langle N \rangle = 1000$)
discourages the simple approach of locating the FM/NF 
phase boundary by seeking where $m$ vanishes, and we instead 
seek a sharper signal for the phase transition, magnetic percolation. 
In graph (b), the $T=0$ magnetic susceptibility $\chi = \left[ \langle {\bf S}^{2}
\rangle \right] - \left[ \langle \left| {\bf S} \right| \rangle \right ]^{2}$ 
(with 
a peak at $n_{c}/n_{i} \sim 0.1$) is shown. 
Similar singular behavior is also a feature of the finite 
temperature second order FM to NF transition (where the temperature is varied
with all other parameters held fixed), and the peak in $\chi(T=0)$ suggests
that our FM to NF phase transition represents
a zero temperature critical point where the critical behavior is driven by 
disorder rather than by thermal fluctuations, a hallmark of a 
percolation transition.

Panel (c) of 
Fig.~\ref{Fig:fig1} depicts the Curie temperatures $T_{c}$ 
as a function of $n_{c}/n_{i}$; for the undamped 
case (filled symbols) the $T_{c}$ curve attains a 
maximum value for $n_{c}/n_{i} \sim 0.03$ and then declines, 
vanishing for $n_{c}/n_{i} \approx 0.1$ in sharp 
contrast with Curie-Weiss continuum mean 
field theory, which yields the monotonically 
increasing $T_{c} \propto n_{i} n_{c}^{1/3}$.
For the damped RKKY model (open symbols) with 
$l=a$, the ferromagnetic transition temperature 
peaks for $n_{c}/n_{i} \sim 0.12$, and is nonzero
over a broader range than for the pure RKKY model.  However,
as can be seen from the graph, the 
highest $T_{c}$ for the damped model is considerably 
suppressed relative to the peak Curie Temperature for the 
$l = \infty$ case.
Panel (d) of Fig.~\ref{Fig:fig1} 
also displays $T_{c}$, but with $n_{c}/n_{i} = 0.03$ 
fixed and the impurity concentration $x_{i}$ allowed to vary.
The open circles are the Monte Carlo results for $l = \infty$, while the 
solid curve is a theoretical curve given by $T^{'} = 29.2x_{i}^{4/3}$.
It can be shown that the good agreement of the $x_{i}^{4/3}$ law 
with the calculated Curie temperatures 
implies that even for $x_{i}$ as large as $0.2$ the system is in the 
dilute limit (i.e. thermodynamic quantities are insensitive to 
the details of the lattice structure).  However, this does not mean
that the dependence of $T_{c}$
is correctly given by continuum MFT, and one actually has
$T_{c} = x_{i}^{4/3}g(n_{c}/n_{i})$, 
where $g$ is constant in continuum Curie-Weiss
theory, but in our case has a nontrivial dependence as highlighted 
in Fig.~\ref{Fig:fig1} (c) and in the inset, which shows
MFT and Monte Carlo $T_{c}$'s on the same graph for $l=a$.

\begin{figure}
\centerline{\psfig{figure=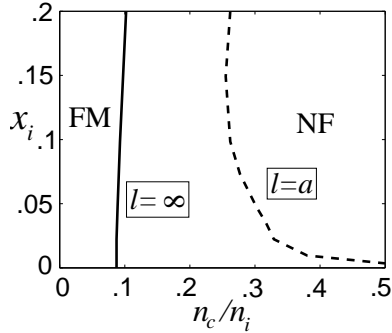,width=2in}}
\vspace{.2cm}
\caption{Phase diagrams for the RKKY model; 
the solid line corresponds to the undamped, full RKKY model while the 
broken line is for the damped case where $l = a$.} 
\label{Fig:fig2}
\end{figure}

By finding the critical
$n_{c}/n_{i}$ values for various Mn concentrations $x_{i}$, we construct 
a phase diagram for the RKKY model for DMS systems, and the 
result is shown in panel (a) of Fig.~\ref{Fig:fig2} for the
undamped ($l = \infty$) case.  From the 
vertical axis, one sees that 
a substantial range of $x_{i}$ values is included, 
certainly encompassing the experimentally relevant range. 
Nonetheless, there is little variation of location of the phase boundary 
which appears essentially as a vertical line at the low (but finite) 
$n_{c}/n_{i} = 0.1$.
This is consistent with the notion that the dilute limit has 
been reached even for Mn doping levels as high as $20 \%$.

As ferromagnetism in 
$\textrm{Ga}_{1-x}\textrm{Mn}_{x}\textrm{As}$ is
found in experiment to be robust for $n_{c}/n_{i} > 0.1$, 
for a more
realistic treatment we consider a damping scale equal to the 
crystal lattice constant, $l=a$.  Since the factor $e^{-r/l}$
attenuates the more distant AF couplings more sharply than the ferromagnetic 
interactions at closer range, one expects RKKY damping to  
extend the FM/NF phase boundary to greater 
$n_{c}/n_{i}$ values, and the broken line in 
Fig.~\ref{Fig:fig2}, which displays the FM/NF phase boundary 
for $n_{c}/n_{i} \geq 0.25$, is consistent with this intuition.
It is important to note that although damping 
expands the domain of the ferromagnetic phase by suppressing  
AF couplings more severely than ferromagnetic interactions, the 
fact that the 
latter are reduced
leaves the ferromagnetic state more readily disrupted by
thermal fluctuations as can be seen in Fig.~\ref{Fig:fig1} where
$T_{c}$ values for the ($l=\infty$) and the ($l=a$) cases are
shown together.  Note that an alternate possibility for an 
extended FM regime in the phase diagram could be the Fermi
surface warping~\cite{cincoii} due to band structure effects in GaMnAs. 
%%%%%%%%%%%%%%%%%%%%%% FIG. 2 %%%%%%%%%%%
%%%%%%%%%%%%%%%%%%%%%%%%%%%%%%%%%%%%%%%%%%%%%%%%%%%%%

\begin{figure}
\centerline{\psfig{figure=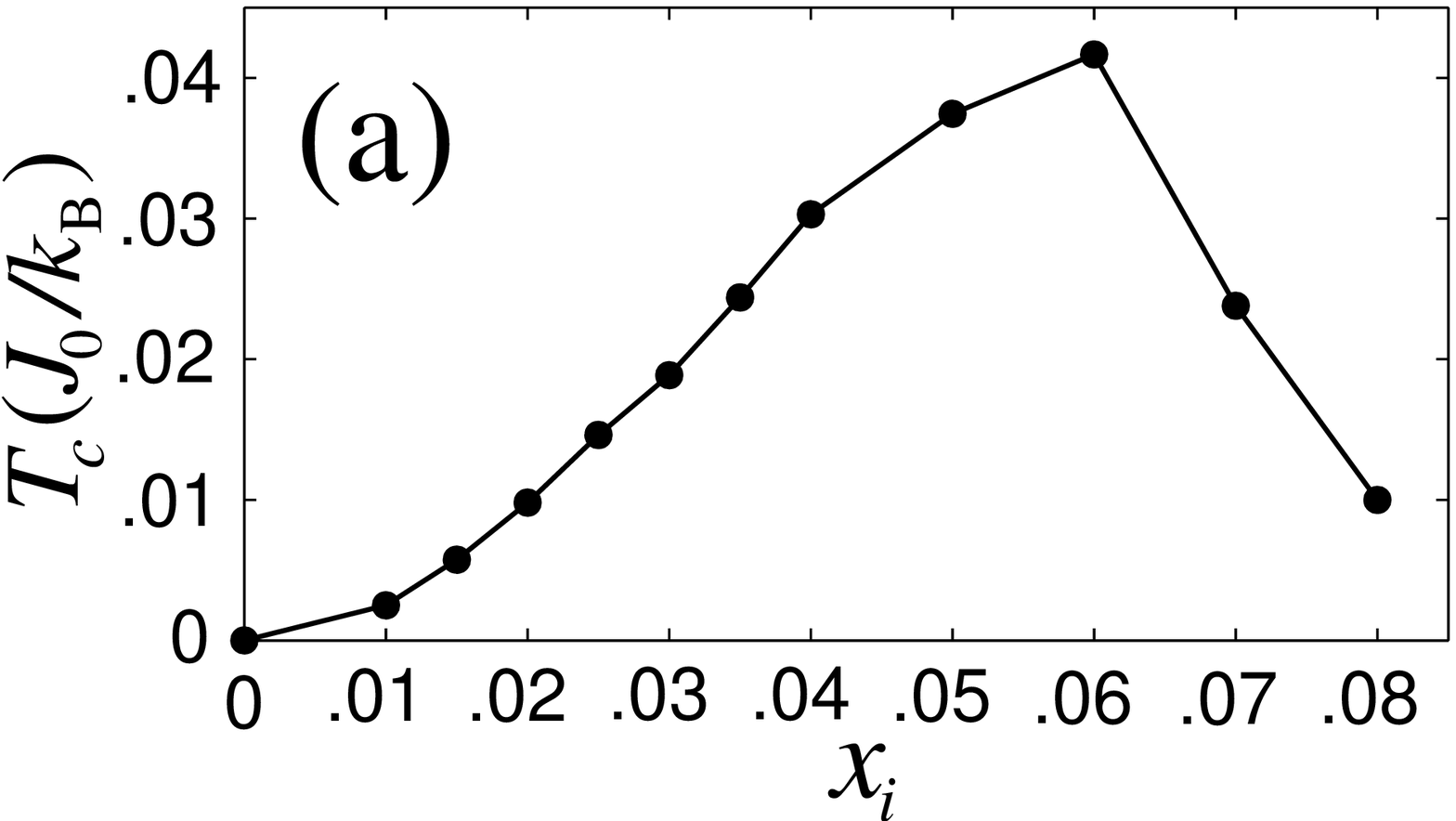,width=2.2in}}
\vspace{.2cm}
\centerline{\psfig{figure=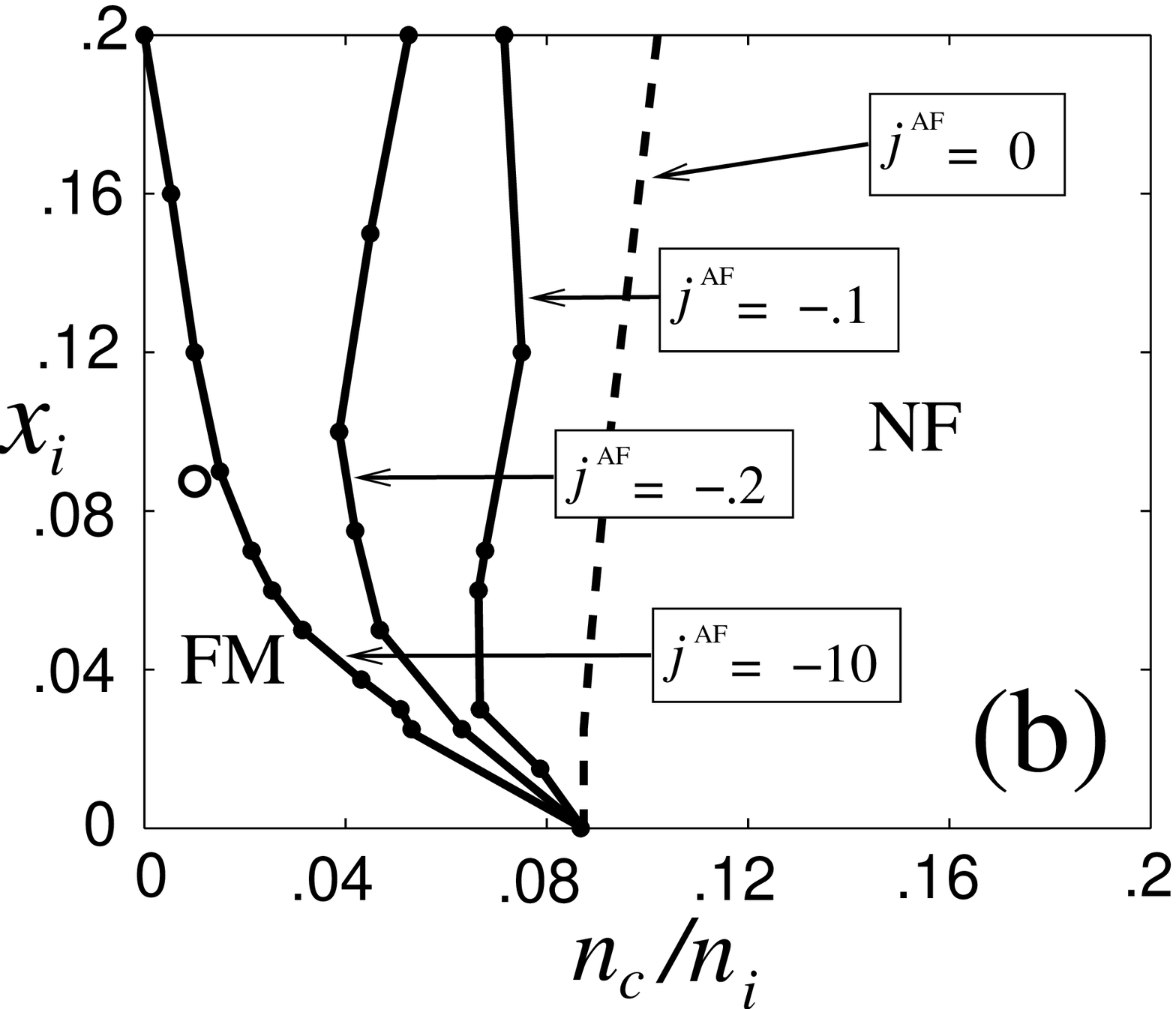,width=2.2in}}
\vspace{.2cm}
\centerline{\psfig{figure=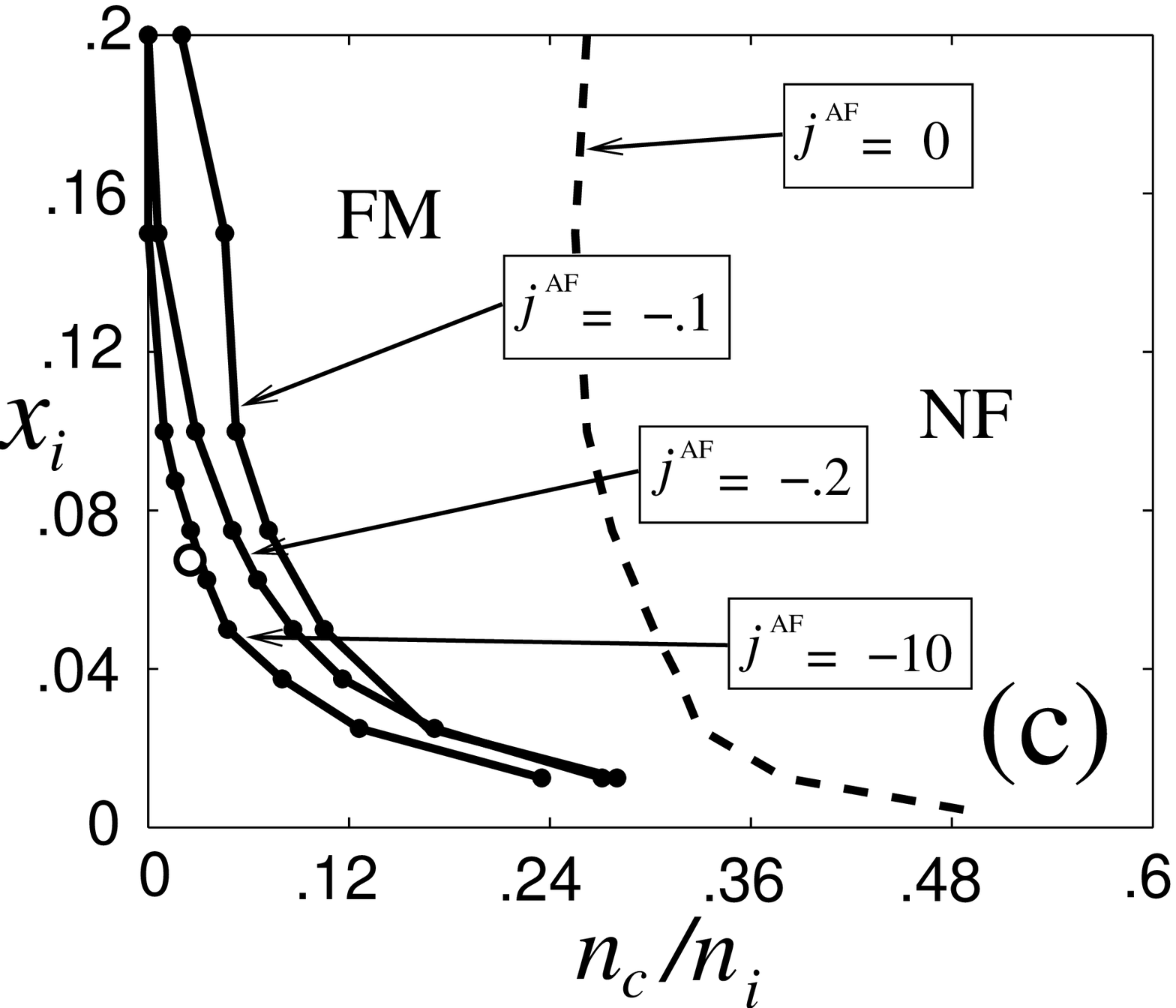,width=2.2in}}
\caption{The graph in panel (a) displays $T_{c}$ versus $x$ for
$n_{c}/n_{i}=0.05$, $l = a$, and $j^{\mathrm{AF}}=-10$.
Panels (b) and (c) are phase diagrams for the RKKY model with
a local AF coupling incorporated for $l = \infty$ (b)
and $l = a$ (c); the open circles at the left
denote locations in the diagrams where
$T_{c}$ is optimized.}
\label{Fig:fig3}
\end{figure}

For $\textrm{Ga}_{1-x}\textrm{Mn}_{x}\textrm{As}$, 
a salient early experimental finding 
is that $T_{c}$ is maximized for 
$x_{i}=x_{i}^{\mathrm{opt}} \sim 0.07$
(though this is dependent on the details of sample preparation)
with the Curie temperature decreasing for doping levels above
or below this optimal concentration.
We consider strong
local superexchange 
AF couplings (i.e. between nearest neighbors on the fcc lattice)
as a major contribution to the weakening
of the ferromagnetic phase for higher impurity concentrations,
where neighboring pairs of Mn impurities are 
more common.  These adjacent moments in the fcc lattice
would presumably form spin singlets and hence not 
contribute to the ferromagnetic phase.  
Since the value of the local Mn-Mn coupling $J^{\mathrm{AF}}$ is 
not precisely known, we examine a representative set of values, and
we work in terms of the rescaled superexchange coupling 
$j^{\mathrm{AF}} \equiv J^{\mathrm{AF}}/J^{\mathrm{RKKY}}_{\mathrm{max}}$,
where $J^{\mathrm{RKKY}}_{\mathrm{max}} = 4 \pi J_{0}$ is the maximum 
possible ferromagnetic coupling possible (for the $l= \infty$ 
case) for 
two nearest neighbor impurities on the fcc lattice.
In Fig.~\ref{Fig:fig3} (a), we show a graph of $T_{c}$  
versus $x$ for a large AF superexchange ($j^{\mathrm{AF}} = -10$).
with $n_{c}/n_{i} = 0.05$ and $l=a$ held fixed; $T_{c}$ peaks
for $x_{i}^{\mathrm{opt}}=.06$, in
reasonable accord with experiment.  In general for any finite
value of $J^{\mathrm{AF}}$, larger $x_{i}$ values will strongly 
suppress the FM phase due to the direct AF coupling 
between the Mn moments.

Using the technique employed for  
$J^{\mathrm{AF}}=0$,
we obtain the $T=0$ phase diagram for both the damped and undamped
cases with results displayed in panels (b) ($l = \infty$) and (c)
($l = a$) of Fig.~\ref{Fig:fig3} for several 
$j^{\mathrm{AF}}$ values.  In both the damped and undamped cases, it is 
evident that a strong to moderate $j^{\mathrm{AF}}$ sharply  
restricts the domain of the ferromagnetic phase; in (b) and (c), 
the ferromagnetic region quickly narrows as the impurity concentration 
rises beyond a few percent, effectively cutting off
ferromagnetism for $x_{i} \gtrsim 0.1$.  Only in the  
dilute limit, where neighboring impurity pairs are less 
abundant do the 
the FM/NF phase boundaries in (b) and (c) revert to 
their position in the $j^{\mathrm{AF}} = 0$ case.
For the case of a very strong local AF interaction
(with $j^{\mathrm{AF}} = -10$), we have determined
the values of $x_{i}$ and $n_{c}/n_{i}$ which maximize $T_{c}$.
The points where $T_{c}$ is optimized are identified
in the phase diagrams of Fig.~\ref{Fig:fig3} (b) and (c)
with open circles at the left of
both panels.
%%%%%%%%%%%%%%%%%%%%%% FIG. 3 %%%%%%%%%%%
%%%%%%%%%%%%%%%%%%%%%%%%%%%%%%%%%%%%%%%%%%

    Finally, we comment on the formation of 
magnetic clusters above $T_{c}$, for the strongly damped case
where the length scale $l$ of $J(r)$ is much smaller 
than the typical separation $l_{s}$ between spins.  
It has been suggested~\cite{ocho} that in this 
limit, one can define a clustering temperature $T^{*}$ above $T_{c}$ where
substantially sized magnetic clusters form.  In fact, 
one can argue that 
$\xi \propto l_{s}(l_{s}/l)^{2/3}[\ln (T/T_{c})]^{-2/3}$,
indicating that the cluster size $\xi$ is only weakly 
dependent on $T/T_{c}$, with much greater sensitivity to 
$l_{s}/l$.  One can also seek a $T^{*}$ below which
typical magnetic clusters are at least $\xi = \alpha l_{s}$
in size, where $\alpha$ is a dimensionless factor, and 
one finds $T^{*} = \exp [ \alpha^{3/2}
(l_{s}/l) ]T_{c}$; it is clear that $T^{*}$ rapidly becomes
large relative to $T_{c}$ in the $l \ll l_{s}$ limit.

In conclusion, we have worked out the $T=0$ phase diagram
of the the full RKKY
model in dilute limit (i.e. in the DMS context) 
at $T=0$, finding that a ferromagnetic
ground state is indeed supported, albeit over a very small 
region of the phase diagram, while continuum mean 
field theory erroneously assumes ferromagnetic order
as the stable $T=0$ phase leading to  
$T_{c}^{\mathrm{MFT}} \propto n_{i} n_{c}^{1/3}$, quite distinct from
our non-monotonic $T_{c}$.
    We have found that the $T=0$ phase diagram of the 
RKKY model consists of 
ferromagnetic and non-ferromagnetic 
phases separated by a line of $T=0$ critical points in which
the NF phase gives way to ferromagnetic order
via the percolation of magnetic clusters.
   We note that a zero temperature transition of a 
NF to a FM phase signaled by 
magnetic percolation is not special to the 
RKKY model in the DMS context, but is of broad relevance 
to magnetic systems where there is strong disorder and
circumstances which allow for competing antiferromagnetic
interactions of a tunable strength.  We have found that
introducing a cutoff in the range of the RKKY 
interaction (arising, for example, from disorder or
carrier localization effects) and including a local AF superexchange term
leads to a $T=0$ phase diagram and $T_{c}$ behavior in reasonable agreement 
with experiment.  
The FM phase of the RKKY DMS model is 
fragile, existing over a restricted parameter space
($l= \infty$) with moderate $T_{c}$'s or on a broader
domain (finite $l$) of parameter space with reduced $T_{c}$'s.

   This work has been supported by US-ONR and LPS.


\begin{thebibliography}{99}

\bibitem{tres} A. Chattopadhyay {\it et al}, Phys. Rev. Lett.
{\bf 87}, 227202 (2001).

\bibitem{cuatro} L. R. Walker {\it et al}, Phys. Rev. B
{\bf 22}, 3816 (1980).

\bibitem{cinco} A. Chakrabarti {\it et al}, Phys. Rev. Lett. 
{\bf 56}, 1404 (1986).

\bibitem{cincoii} L. Brey {\it et al}, Phys. Rev. B
{\bf 68}, 115206 (2003).

\bibitem{cincoiii} C. G. Zhou {\it et al}, Phys. Rev. B
{\bf 69}, 144419 (2004).

\bibitem{cincoiv} J. Schliemann {\it et al}, Phys. Rev. B
{\bf 64}, 165201 (2001).

\bibitem{cincovi} A. Kaminski {\it et al}, Phys. Rev. Lett.
{\bf 88}, 247202 (2002).

\bibitem{seis} J. K. Kim, Phys. Rev. D {\bf 50}, 4663 (1994).

\bibitem{siete} Y. Miyatake {\it et al}., J. Phys. C {\bf 19}
2539 (1986).

\bibitem{ocho} G. Alvarez {\it et al}, Phys. Rev. Lett.
{\bf 89}, 277202 (2002).

\end{thebibliography}
\end{document}